\documentclass{llncs}
\usepackage{llncsdoc}
\usepackage{amsmath}
\usepackage{datetime}
\usepackage{amsfonts}
\usepackage{textcomp}
\usepackage{algorithmic}
\usepackage{amssymb}
\usepackage{layout}
\usepackage{algorithm}
\usepackage{latexsym}
\usepackage{color}
\usepackage{mathrsfs}
\usepackage{subfigure}
\usepackage{graphicx}
\usepackage{epsfig}
\usepackage{tikz}
\usepackage{enumerate}
\usetikzlibrary{arrows,calc,shapes, snakes, intersections,patterns}

\newcommand{\gettikzxy}[3]{%
  \tikz@scan@one@point\pgfutil@firstofone#1\relax
  \edef#2{\the\pgf@x}%
  \edef#3{\the\pgf@y}%
}

\newcommand{\A}{\mathcal{A}}
\newcommand{\F}{\mathcal{F}}
\newcommand{\T}{\mathcal{T}}
\newcommand{\E}{\mathbb{E}}
\newcommand{\C}{\mathcal{R}}
\newcommand{\y}{\textbf{b$^{\prime}$}}
\newcommand{\z}{\textbf{b$^{\prime\prime}$}}

\newenvironment{pfof}{\noindent{\em Proof:} }{ \hfill $\blacksquare$\\ }

\begin{document}

\newcounter{save}\setcounter{save}{\value{section}}

\title{Profit Maximizing Prior-free Multi-unit \\Procurement Auctions with Capacitated Sellers}
\author{Arupratan Ray\inst{1}, Debmalya Mandal\inst{2}, \and Y. Narahari\inst{1}}
\institute{Department of Computer Science and Automation, Indian Institute of Science, Bangalore, India, \email{rayarupratan@gmail.com, hari@csa.iisc.ernet.in} \and Harvard School of Engineering and Applied Sciences, Cambridge, MA, \email{dmandal@g.harvard.edu}}

\maketitle
\bibliographystyle{splncs}

\begin{abstract}
In this paper, we derive bounds for profit maximizing prior-free
procurement auctions where a buyer wishes to procure multiple units of a
homogeneous item from $n$ sellers who are strategic about their per unit valuation. 
The buyer earns the profit by
reselling these units in an external consumer market. 
The paper looks at three scenarios of increasing complexity : 
(1) sellers with unit capacities, 
(2) sellers with non-unit capacities which are common knowledge, 
and (3) sellers with non-unit capacities which are private to the sellers.
First, we look at unit capacity sellers where per unit valuation is private information
of each seller and the revenue curve is concave. For this setting, we define two benchmarks. We show that
no randomized prior free auction can be constant competitive against any of these two
benchmarks. However, for a lightly constrained benchmark we design
a prior-free auction PEPA (Profit Extracting Procurement Auction) which is 4-competitive and we show this bound is tight.
Second, we study a setting where the sellers have
non-unit capacities that are common knowledge and derive similar results. In particular,
we propose a prior free auction PEPAC (Profit Extracting Procurement Auction with Capacity)
which is truthful for any concave revenue curve.
Third, we obtain results in the inherently harder bi-dimensional case
where per unit valuation as well as capacities are private
information of the sellers. We show that PEPAC is truthful and constant competitive for the specific case of
linear revenue curves. 
We believe that this paper represents the first set of results on 
single dimensional and bi-dimensional 
profit maximizing prior-free multi-unit procurement auctions.
\end{abstract}

\section{Introduction}

Procurement auctions for awarding contracts to supply goods or services are prevalent in many modern resource
allocation situations. In several of these 
scenarios, the buyer plays the role of an  intermediary who purchases some goods or services from 
the suppliers and resells it in the 
consumer market. For example, in the retail sector, an intermediary procures products from different vendors
(perhaps through an auction)
and resells it in  consumer markets for a profit. 
In a cloud computing setting \cite{abhinandan2013}, an intermediary buys cloud resources
from different service providers (again through an auction) and resells these resources to
requesters of cloud services. 
The objective of the intermediaries in each of these cases is to maximize the profit earned
in the process of reselling.

Solving such problems via optimal auction of the kind discussed in the auction literature \cite{myerson1981optimal} inevitably
requires assumption of a \emph{prior} distribution on the sellers' valuations.  
The requirement of a known prior distribution often 
places severe practical limitations.  We have to be extremely careful in using a prior distribution which is collected from 
the past transactions of the bidders as they can possibly manipulate to do something differently than before. 
Moreover, deciding the prior distribution ideally requires a large number of samples. In reality, we can only approximate it with a finite number of 
samples. Also, prior-dependent auctions are non-robust: if the prior distribution alters, 
we are compelled to repeat the entire computation for the new 
prior, which is often  computationally hard. This motivates us to study \emph{prior-free} auctions. In particular, 
in this paper, we study profit maximizing
prior-free procurement auctions with one buyer and $n$ sellers. 

\section{Prior Art and Contributions}
The problem of designing a revenue-optimal auction was first studied by Myerson \cite{myerson1981optimal}. 
Myerson considers the setting of a seller trying to sell a single object to one of several possible buyers 
and characterizes all revenue-optimal auctions that are BIC (Bayesian Incentive Compatible) and IIR
(Interim individually Rational). 
Dasgupta and Spulber \cite{dasgupta1990managing} 
consider the problem of designing an optimal procurement auction where suppliers have unlimited capacity. Iyengar and Kumar \cite{iyengar2008optimal} consider the setting
where the buyer purchases multiple units of a single item from the suppliers and resells it in the consumer market to 
earn some profit. We consider the same setting here, however, we focus on the design of \emph{prior-free} auctions
unlike the \emph{prior-dependent} optimal auction designed in \cite{iyengar2008optimal}. 

\subsection{Related Work and Research Gaps}
Goldberg et al. \cite{goldberg2001competitive} initiated work on design of prior-free auctions and studied a class of single-round 
sealed-bid auctions for an item in unlimited supply, such as digital goods where each bidder requires at most one unit. 
They introduced the notion of \emph{competitive} auctions and proposed prior-free randomized competitive 
auctions based on random sampling. In \cite{goldberg2006competitive}, the authors consider non-asymptotic behavior 
of the random sampling optimal price auction (RSOP) and show that its performance is within a large constant factor of a prior-free 
benchmark. Alaei et al. \cite{alaei2009random} provide a nearly tight analysis of RSOP which shows that it is 4-competitive for a large class of 
instances and 4.68 competitive for a much smaller class of remaining instances. The competitive ratio has further been improved to 3.25 
by Hartline and McGrew \cite{hartline2005optimal} and to 3.12 by Ichiba and Iwama \cite{ichiba2010averaging}. 
Recently, Chen et al. \cite{chen2014optimal} designed an optimal competitive auction where the competitive ratio matches 
the lower bound of 2.42 derived in \cite{goldberg2006competitive} and therefore settles an important open problem in the design of digital 
goods auctions.

Beyond the digital
goods setting, Devanur et al. \cite{devanur2012envy} have studied prior-free auctions under several settings
such as multi-unit, position auctions, downward-closed, and matroid environments. They design a prior-free auction with competitive 
ratio of 6.24 for the multi-unit and position auctions using the 3.12 competitive auction for digital goods given in \cite{ichiba2010averaging}. 
They also design an auction with a competitive ratio of 12.5 for multi-unit settings by directly generalizing 
the random sampling auction given in \cite{goldberg2006competitive}. 
Our setting is different than the above works on forward auction as we consider a procurement auction with capacitated sellers.

Somewhat conceptually closer to our work is budget feasible mechanism design 
(\cite{singer2010budget}, \cite{bei2012budget}, \cite{chen2011approximability}, \cite{dobzinski2011mechanisms}) 
which models a simple procurement auction. Singer \cite{singer2010budget} 
considers single-dimensional mechanisms that maximize the buyer's valuation function on subsets of items, under the constraint that the sum of the 
payments provided by the mechanism does not exceed a given budget. Here the objective is maximizing social welfare derived 
from the subset of items procured under a budget and the benchmark considered is welfare optimal. On the other hand, our work considers maximizing profit or revenue of the 
buyer which is fundamentally different from the previous objective and our benchmark is revenue optimal. 
A simple example can be constructed to show that Singer's benchmark is not revenue optimal as follows. Suppose  
there is a buyer with budget \$100 and valuation function $V(k) = \$25k$ ($k$ is the number of items procured) and $5$ sellers with costs \$10, \$20, \$50, \$60, and \$70. 
Then Singer's benchmark will procure 3 items (with costs \$10, \$20, and \$50) earning a negative utility to the buyer which is welfare 
optimal but not revenue optimal as an omniscient revenue maximizing allocation will procure 2 items (with costs \$10 and \$20) yielding a 
revenue or utility of \$20 to the buyer.

Although the design of prior-free auctions has generated wide interest in the research community 
(\cite{alaei2009random}, \cite{chen2014optimal}, \cite{devanur2013prior}, 
\cite{devanur2012envy}, \cite{goldberg2006competitive}, \cite{goldberg2001competitive}, \cite{hartline2005optimal}, 
\cite{ichiba2010averaging})
most of the works have considered the forward setting. 
The reverse auction setting is subtly different from
forward auctions especially if the sellers are capacitated and
the techniques used for forward auctions cannot be trivially extended
to the case of procurement auctions.
To the best of our knowledge, design of profit-maximizing prior-free multi-unit procurement auctions is yet unexplored. 

Moreover, the existing literature on 
prior-free auctions is limited to the single-dimensional setting where each bidder has only one private type which is valuation per unit of an item. 
However, in a procurement auction, the sellers are often capacitated and strategically report their capacities to increase 
their utilities. Therefore, the design of bi-dimensional prior-free procurement auctions is extremely relevant in practice and in 
this paper, we believe we have derived the first set of results in this direction. 

\subsection{Contributions}
In this paper, we design profit-maximizing prior free procurement auctions  where a buyer procures multiple units of an item 
from $n$ sellers and subsequently resells the units to earn revenue. Our contributions are three-fold. 
First, we look at unit capacity sellers and define two benchmarks for analyzing the performance of any prior-free 
auction -- (1) an optimal single price auction $(\F)$ and 
(2) an optimal multi-price auction $(\T)$. We  show that no prior-free auction can be 
constant competitive against any of the two benchmarks. 
We then consider a lightly relaxed benchmark $(\F^{(2)})$ which is constrained to procure at least two units and design a prior-free auction 
PEPA (Profit Extracting Procurement Auction) 
which is 4-competitive against $(\F^{(2)})$ for any concave revenue curve. 
Second, we study a setting where the sellers have
non-unit capacities that are common knowledge and derive similar results. In particular,
we propose a prior free auction PEPAC (Profit Extracting Procurement Auction with Capacity)
which is truthful for any concave revenue curve.
Third, we obtain results in the inherently harder bi-dimensional case
where per unit valuation as well as capacities are private
information of the sellers. We show that PEPAC is truthful and constant competitive for the specific case of  
linear revenue curves. 

We believe the proposed auctions represent the first effort in single dimensional and bi-dimensional prior-free multi-unit procurement auctions. 
Further, these auctions can be easily adapted in real-life procurement situations due to the simple rules and prior-independence of the auctions.

\section{Sellers with Unit Capacities}
\label{model1}
We consider a single round procurement auction setting with one buyer (retailer, intermediary etc.) and $n$ sellers where each seller 
has a single unit of a homogeneous item. The buyer procures multiple units of the item from the 
sellers and subsequently resells it in an outside consumer market, earning a revenue of $\C(q)$ from selling $q$ units of the item. 
We assume that the revenue curve of the outside market $\C(q)$ is concave with $\C(0) = 0$. This is motivated by the following standard 
argument from economics. According to the \emph{law of demand}, the quantity demanded decreases as price per unit increases. It can be 
easily shown that the marginal revenue falls with an increase in the number of units sold and so the revenue curve is concave.

\subsection{Procurement Auction}
We assume that the buyer (auctioneer) has \emph{unlimited demand}, but as each seller (bidder) has unit capacity, the number of units the buyer can procure and resell is limited by the total number of sellers. We make the following assumptions about the bidders:
\begin{enumerate}
 \item Each bidder has a private valuation $v_i$ which represents the true minimum amount he is willing to receive to sell a single unit of the item.
 \item Bidders' valuations are independently and identically distributed.
 \item Utility of a bidder is given as payment minus valuation.
\end{enumerate}

In reference to the \emph{revelation principle} 
\cite{myerson1981optimal}, we will restrict our attention to single-round, sealed-bid, truthful auctions. 
Now we define the notions of single-round sealed-bid auction, bid-independent auction, and competitive auction 
for our setting. 

\subsubsection*{Single-round Sealed-bid Auction ($\A$).}
 \begin{enumerate}
  \item The bid submitted by bidder $i$ is $b_i$. The vector of all the submitted bids is denoted by $\textbf{b}$. Let \textbf{b}$_{-i}$ denote the masked vector of bids where $b_i$ is removed from \textbf{b}.
  \item Given the bid vector $\textbf{b}$ and the revenue curve $\C$, the auctioneer computes an allocation  $\textbf{x} = 
  (x_1, \ldots, x_n)$, and payments $\textbf{p} = (p_1, \ldots, p_n)$. If bidder $i$ sells the item, $x_i = 1$ and we say 
  bidder $i$ wins. Otherwise, bidder $i$ loses and $x_i = 0$. The auctioneer pays an amount $p_i$ to bidder $i$. We assume that 
  $p_i \geq b_i$ for all winning bidders and $p_i = 0$ for all losing bidders.
  \item The auctioneer resells the units bought from the sellers in the outside consumer market. The profit of the auction 
  (or auctioneer) is given by,
  \begin{center}
   $\A(\textbf{b}, \C) = \C(\sum_{i=1}^{n}{x_i(\textbf{b}, \C)}) - \sum_{i=1}^{n}{p_i(\textbf{b}, \C)}$.
  \end{center}
 \end{enumerate} 

 The auctioneer wishes to maximize her profit satisfying IR (Individual Rationality) and DSIC (Dominant Strategy Incentive Compatibility). 
 As bidding $v_i$ is a dominant strategy for bidder $i$ in a 
 truthful auction, in the remainder of this paper, we assume that $b_i = v_i$.
 
\subsubsection*{Bid-independent Auction.}
An auction is \emph{bid-independent} if the offered payment to a bidder is independent of the bidder's bid. It can certainly depend on 
the bids of the other bidders and the revenue curve. Such an auction is determined by a function $f$ (possibly randomized) which takes the masked bid vectors and 
the revenue curve as input and maps it to payments which are non-negative real numbers. Let $\A_f($\textbf{b}$, \C)$ denote the 
bid-independent auction defined by $f$. For each bidder $i$ the allocation and payments are determined in two phases as follows :
 \begin{enumerate} 
 \item Phase I :
 \begin{enumerate}
  \item $t_i \leftarrow f(\textbf{b}_{-i}, \C)$. 
  \item If $t_i < b_i$, set $x_i \leftarrow 0$, $p_i \leftarrow 0$, and remove bidder $i$.
 \end{enumerate} 
 
 Suppose $n^\prime$ is the number of bidders left. Let $t_{[i]}$ denote the $i^{th}$ lowest value of $t_j$ among the remaining $n^\prime$ bidders.
 Let $x_{[i]}$ and $p_{[i]}$ be the corresponding allocation and payment. 
 Now we choose the allocation that maximizes the revenue of the buyer. 
 \item Phase II : 
 \begin{enumerate}
  \item $k \leftarrow \underset{0 \leq i \leq n^\prime}{\operatorname{argmax}}\  (\C(i) - \sum_{j=1}^i{t_{[j]}})$.
  \item Set $x_{[i]} \leftarrow 1$ and $p_{[i]} \leftarrow t_{[i]}$ for $i = \{1, \ldots, k\}$.
  \item Otherwise, set $x_{[i]} \leftarrow 0,  p_{[i]} \leftarrow 0$.
 \end{enumerate}
\end{enumerate} 

For any bid-independent auction, the allocation of bidder $i$ is non-increasing in valuation $v_i$ and his payment is independent of his bid. 
It follows from Myerson's characterization \cite{myerson1981optimal} of truthful auctions that any bid-independent auction is truthful.

\subsubsection*{Competitive Auction.}
In the absence of any prior distribution over the valuations of the bidders, we cannot compare the profit of a procurement auction with respect to the average profit of the optimal auction. Rather, we measure the performance of a truthful auction on any bid by comparing it with the profit that would have been achieved 
by an omniscient optimal auction ($OPT$), the optimal auction which knows all the true valuations in advance without requiring to elicit them from the bidders. 

\begin{definition}
 \emph{$\beta$-competitive auction ($\beta > 1$)} :
 An auction $\A$ is \emph{$\beta$-competitive} against $OPT$ if for all bid vectors \textbf{b}, the expected profit of $\A$ on $\textbf{b}$ 
 satisfies
 \begin{center}
  $\E[\A(\textbf{b}, \C)] \geq \displaystyle \frac{OPT(\textbf{b}, \C)}{\beta}$.
 \end{center}
 We refer to $\beta$ as the \emph{competitive ratio} of $\A$.
 Auction $\A$ is \emph{competitive} if its competitive ratio $\beta$ is a constant.
\end{definition}

\subsection{Prior-Free Benchmarks}
\label{bench}
As a first step in comparing the performance of any prior-free procurement auction, we need to come up with the right metric
for comparison that is a benchmark. It is important that we choose such a benchmark carefully for such a comparison to be 
meaningful. Here, we start with the strongest possible benchmark for comparison: the profit of an auctioneer who knows the bidder's true valuations. This leads us to consider the two most natural metrics for comparison -- the optimal multiple price and single price 
auctions. We compare the performances of truthful auctions to that of the optimal multiple price and single price auctions. 
Let $v_{[i]}$ denote the $i$-th lowest valuation.

\subsubsection*{Optimal Single Price Auction ($\F$).}
 Let $\textbf{b}$ be a bid vector. Auction $\F$ on input $\textbf{b}$ determines the value $k$ such that $\C(k) - kv_{[k]}$ is 
 maximized. All bidders with bid $b_i \leq v_{[k]}$ win at price $v_{[k]}$; all remaining bidders lose. We denote the optimal 
 procurement price for \textbf{b} that gives the optimal profit by $OPP(\textbf{b}, \C)$. 
 The profit of $\F$ on input $\textbf{b}$ is denoted by $\F(\textbf{b}, \C)$. So we have, 
 \begin{center}
  $\F(\textbf{b}, \C) = \max \limits_{0 \leq i \leq n} (\C(i) - iv_{[i]})$.\\
  $OPP(\textbf{b}, \C) = \underset{v_{[i]}}{\operatorname{argmax}}\  (\C(i) - iv_{[i]})$.
 \end{center} 
 
\subsubsection*{Optimal Multiple Price Auction ($\T$).}
 Auction $\T$ buys from each bidder at her bid value. So auction $\T$ on input $\textbf{b}$ determines the value $l$ such that 
 $\C(l) - \sum_{i=1}^{l}{v_{[i]}}$ is maximized. First $l$ bidders win at their bid value; all remaining bidders lose. 
 The profit of $\T$ on input $\textbf{b}$ is given by
 \begin{center}
  $\T(\textbf{b}, \C) = \max \limits_{0 \leq i \leq n} (\C(i) - \sum_{j=1}^{i}{v_{[j]}})$.
 \end{center}

It is clear that $\T(\textbf{b}, \C) \geq \F(\textbf{b}, \C)$ for any bid vector 
$\textbf{b}$ and any revenue curve $\C$. However, $\F$ does not perform very poorly compared to $\T$. 
We prove a bound between the performance of $\F$ and $\T$. 
Specifically, we observe that in the worst case, the maximum ratio of $\T$ to $\F$ is logarithmic in the number $n$ of bidders.

\begin{lemma}
 \label{benchmarkrelation}
 For any \normalfont \textbf{b} \textit{and any concave revenue curve $\C$},  
 \begin{center}
 $\F(\textbf{b}, \C) \geq \displaystyle \frac{\T(\textbf{b}, \C)}{\mathrm{ln} \ n}$.
\end{center}
\end{lemma}
\begin{pfof}
 We use the following property of concave function, $\displaystyle \frac{\C(i)}{i} \geq \frac{\C(j)}{j}$  $\ \ \forall i \leq j$.
 Suppose $\T$ buys $k$ units and $\F$ buys $l$ units from the sellers.
 \begin{align*}
  \T(\textbf{b}, \C) &= \C(k) - \sum_{i=1}^{k}{v_{[i]}} = \sum_{i=1}^{k}\left({\frac{\C(k)}{k} - v_{[i]}}\right) \\
  &\leq \sum_{i=1}^{k}\left({\frac{\C(i)}{i} - v_{[i]}}\right)  
  \leq \sum_{i=1}^{k}{\frac{\C(l)- lv_{[l]}}{i}} \\ 
  &\leq \F(\textbf{b}, \C)(\mathrm{ln}\ n + O(1)).
 \end{align*} 
\end{pfof}
The result implies that if an auction $\A$ is constant-competitive against $\F$ then it is $\ln n$ competitive against $\T$. 
Now we show that no truthful auction can be constant-competitive against $\F$ and hence it cannot be competitive against $\T$.
\begin{theorem}
 \label{impossibility}
 For any truthful auction $\A_f$, any revenue curve $\C$, and any $\beta \geq 1$, there exists a bid vector \normalfont \textbf{b} 
 \textit{such that the expected profit of $\A_f$ on} \normalfont \textbf{b} \textit{is at most} $\F($\textbf{b}$, \C)/\beta$.
\end{theorem}
 \begin{pfof}
  Consider a bid-independent randomized auction $\A_f$ on two bids, $r$ and $L < r$ where $r = \C(1)$. 
  Suppose $g$ and $G$ denote the probability density function and cumulative density function of the random variable $f(r, \C)$.
  We fix one bid at $r$ and choose $L$ depending on the two cases.
  \begin{enumerate}
   \item Case I : $G(r) \leq 1/\beta$. We choose $L = \displaystyle \frac{r}{\beta}$. \\
   Then $\F(\textbf{b}, \C) = r \left (1 - \displaystyle \frac{1}{\beta} \right)$ and
   $\E[\A_f(\textbf{b}, \C)] \leq \displaystyle \frac{1}{\beta}\left( r - \frac{r}{\beta} \right) = \frac{\F(\textbf{b}, \C)}{\beta}$.
   \item Case II : $G(r) > 1/\beta$. We choose $L = \displaystyle r - \epsilon$  such that $G(r) - G(r - \epsilon) < 1/\beta$. 
   As $G$ is a non-decreasing function and $G(r) > 1/\beta$ such a value of $\epsilon$ always exists. 
   Then $\F(\textbf{b}, \C) = r - (r - \epsilon) = \epsilon$ and 
   \begin{align*}
    \E[\A_f(\textbf{b}, \C)] &= \displaystyle \int_{r - \epsilon}^{r} (r - y) g(y) \mathrm{d}y\\
    &\leq  \displaystyle r \int_{r - \epsilon}^{r} g(y) \mathrm{d}y - \displaystyle (r - \epsilon)\int_{r - \epsilon}^{r}g(y) \mathrm{d}y
    \\ &= \displaystyle \epsilon \int_{r - \epsilon}^{r} g(y) \mathrm{d}y
    = \displaystyle \epsilon (G(r) - G(r - \epsilon)) \\ &< \frac{\epsilon}{\beta} = \frac{\F(\textbf{b}, \C)}{\beta}. 
   \end{align*}
  \end{enumerate}
\end{pfof}
Theorem \ref{impossibility} shows that we cannot match the performance of the optimal single price auction when 
the optimal profit is generated from the single lowest bid. Therefore we present an 
auction that is \emph{competitive} against $\F^{(2)}$, the optimal single price auction that buys at least two units. Such an
auction achieves a constant fraction of the revenue of $\F^{(2)}$ on \emph{all inputs}.

\subsubsection*{An Auction $\F$ that Procures at least Two Units ($\F^{(2)}$).}
 Let $\textbf{b}$ be a bid vector. Auction $\F^{(2)}$ on input $\textbf{b}$ determines the value $k \geq 2$ such that $\C(k) - kv_{[k]}$ is 
 maximized. The profit of $\F^{(2)}$ on input $\textbf{b}$ is 
 \begin{center}
  $\F^{(2)}(\textbf{b}, \C) = \max \limits_{2 \leq i \leq n} (\C(i) - iv_{[i]})$.
 \end{center}
 Note that, though $\F^{(2)}$ is slightly constrained the performance of $\F^{(2)}$ can be 
 arbitrarily bad in comparison to $\F$. We demonstrate it using a simple example where $\F$ procures only one unit as follows.
\begin{example}
  Consider the revenue curve $\C(k) = rk$ ($r>0$). Let $0 < \epsilon \ll r$ and bid vector 
 $\textbf{b} = (\epsilon, r - \epsilon, r, \ldots, r)$.   
 Then, ${\F(\textbf{b}, \C)} = r - \epsilon$ and $\F^{(2)}(\textbf{b}, \C) = 2r - 2(r - \epsilon) = 2\epsilon$. Hence, 
 $\displaystyle \frac{\F(\textbf{b}, \C)}{\F^{(2)}(\textbf{b}, \C)} = \frac{r - \epsilon}{2\epsilon} = \left(\frac{r}{2\epsilon} - \frac{1}{2}\right)
 \rightarrow \infty$ as $\epsilon \rightarrow 0$. 
 \end{example}
 But if $\F$ chooses to buy at least two units, we have $\F^{(2)}(\textbf{b}, \C) = \F(\textbf{b}, \C)$. Thus, comparing auction performance 
to $\F^{(2)}$ is identical to comparing it to $\F$ if we exclude the bid vectors where only the lowest bidder wins in the optimal 
auction. From now on, we say an auction is \emph{$\beta$-competitive} if it is \emph{$\beta$-competitive} against $\F^{(2)}$. 
\subsection{Profit Extracting Procurement Auction (PEPA)}
\label{rspa}
We now present a prior-free procurement auction based on random sampling. Our auction takes the bids from the bidders and then 
partitions them into two sets by flipping a fair coin for each bid to decide to which partition to assign it. Then we 
use one partition for market analysis and utilize what we learn from a sub-auction on the other partition, and 
vice versa. We extend the \emph{profit extraction} technique of \cite{goldberg2006competitive}. The goal of the technique is, 
given \textbf{b}, $\C$, and profit $P$, to find a subset of bidders who generate profit $P$.
\subsubsection*{Profit Extraction (PE$_P(\textbf{b}, \C)$).}
 Given target profit $P$, 
 \begin{enumerate}
  \item Find the largest value of $k$ for which $v_{[k]}$ is at most $(\C(k) - P)/k$.
  \item Pay these $k$ bidders $(\C(k) - P)/k$ and reject others.
 \end{enumerate} 
\subsubsection*{Profit Extracting Procurement Auction (PEPA).} 
 \begin{enumerate}
  \item Partition the bids \textbf{b} uniformly at random into two sets \textbf{b$^{\prime}$} and \textbf{b$^{\prime\prime}$}: for each bid, flip a fair coin, and with probability 
  1/2 put the bid in \textbf{b$^{\prime}$} and otherwise in \textbf{b$^{\prime\prime}$}.
  \item Compute $F^\prime = \F(\textbf{b}^\prime, \C)$ and $F^{\prime\prime} = \F(\textbf{b}^{\prime\prime}, \C)$ which are the 
  optimal single price profits for \textbf{b$^{\prime}$} and \textbf{b$^{\prime\prime}$} respectively.
  \item Compute the auction results of PE$_{F^{\prime\prime}}$(\textbf{b$^{\prime}$}, $\C)$ and PE$_{F^\prime}$(\textbf{b$^{\prime\prime}$}, $\C)$.
  \item Run the auction PE$_{F^{\prime\prime}}$(\textbf{b$^{\prime}$}, $\C)$ or 
  PE$_{F^\prime}$(\textbf{b$^{\prime\prime}$}, $\C)$ that gives higher profit to the buyer. Ties are broken arbitrarily.   
 \end{enumerate}
The following lemmas can be easily derived.
 \begin{lemma}
  \label{pepa1}
  PEPA \textit{is truthful}.
 \end{lemma}
 \begin{lemma}
  \label{pepa2}
  PEPA \textit{has profit} $F^\prime$ \textit{if} $F^\prime = F^{\prime\prime}$; \textit{otherwise it 
  has profit} $\min(F^\prime, F^{\prime\prime})$.
 \end{lemma}
Now we derive the competitive ratio for PEPA, first for linear revenue curve and then for any arbitrary concave revenue curve.
\begin{theorem}
\label{thm1}
 PEPA is $4$-competitive if the revenue curve is linear i.e. $\C(k) = rk$ where $r > 0$ and this bound is tight.
\end{theorem}
\begin{pfof}
 By definition, $\F^{(2)}$ on \textbf{b} buys from $k \geq 2$ bidders for a profit of $\F^{(2)}(\textbf{b}, \C) = \C(k) - kv_{[k]}$.
 These $k$ bidders are divided uniformly at random between \y \ and \z. Let $k_1$ be the number of them in \y \ and $k_2$ 
 the number in \z. Now we denote the $i^{th}$ lowest bid in \y \ by $v_{[i]}^\prime$ and in \z \ by $v_{[i]}^{\prime\prime}$. Clearly, 
 $v_{[k_1]}^\prime \leq v_{[k]}$ and $v_{[k_2]}^{\prime\prime} \leq v_{[k]}$.
 So we have, $\F^\prime \geq \C(k_1) - k_1v_{[k_1]}^\prime \geq \C(k_1) - k_1v_{[k]}$ and 
 $\F^{\prime\prime} \geq \C(k_2) - k_2v_{[k_2]}^{\prime\prime} \geq \C(k_2) - k_2v_{[k]}$.
 \begin{align*}
  \displaystyle \frac{\min(F^\prime, F^{\prime\prime})}{\F^{(2)}(\textbf{b}, \C)} &\geq \frac{\min(\C(k_1) - k_1v_{[k]}, \C(k_2) - k_2v_{[k]})}{\C(k) - kv_{[k]}} \\
  &= \frac{\min(rk_1 - k_1v_{[k]}, rk_2 - k_2v_{[k]})}{rk - kv_{[k]}} \\ &=  \frac{\min(k_1, k_2)}{k}.
 \end{align*} 
 Thus, the competitive ratio is given by 
 \begin{align}
  \displaystyle \frac{\E[P]}{\F^{(2)}} &= \frac{1}{k} \sum_{i=1}^{k-1}{\min(i, k-i){k \choose i} 2^{-k}}
  = \displaystyle \frac{1}{2} - {k-1 \choose \lfloor{k/2}\rfloor} 2^{-k}.
 \end{align}  
 The above expression achieves its minimum of 1/4 for $k$ = 2 and $k$ = 3. As $k$ increases it approaches 1/2. 
 \vspace{-2mm}
 \end{pfof} 
 \begin{example}
 The bound presented on the competitive ratio is tight. Consider the revenue curve 
 $\C(k) = 2lk$ ($l>0$) and bid vector \textbf{b} which consists of two bids $l - \epsilon$ and $l$. All other bids are very high compared to $l$.
 Then, ${\F(\textbf{b}, \C)} = \F^{(2)}(\textbf{b}, \C) = 2l$. The expected profit of PEPA is $l \cdot \mathbb{P}$ [two low bids are split 
 between \y \ and \z] = $l/2$ = $\F(\textbf{b}, \C)/4$.
 \end{example} 
 \begin{theorem}
 For any concave revenue curve, PEPA is $4$-competitive.
 \end{theorem}
 \begin{pfof}
 Using notation defined above,
 \begin{align*}
 \displaystyle \frac{\min(F^\prime, F^{\prime\prime})}{\F^{(2)}(\textbf{b}, \C)}
  &\geq \frac{\min(\C(k_1) - k_1v_{[k]}, \C(k_2) - k_2v_{[k]})}{\C(k) - kv_{[k]}}\\ 
  &= \frac{\min(k_1(\frac{\C(k_1)}{k_1} - v_{[k]}), k_2(\frac{\C(k_2)}{k_2} - v_{[k]}))}{k(\frac{\C(k)}{k} - v_{[k]})} \\
  &\geq \frac{\min(k_1(\frac{\C(k)}{k} - v_{[k]}), k_2(\frac{\C(k)}{k} - v_{[k]}))}{k(\frac{\C(k)}{k} - v_{[k]})} \\
  &=  \frac{\min(k_1, k_2)}{k} \geq 1/4.
 \end{align*}
 \vspace{-5mm}
 \end{pfof} 
 
\section{Sellers with Non-Unit Non-Strategic Capacities}
\label{Nonstrategic}
\subsection{Setup}
Now we consider the setting where sellers can supply more than one unit of an item. Seller $i$ has valuation per unit $v_i$ 
and a maximum capacity $q_i$ where $v_i$ is a positive real number and $q_i$ is a positive integer. In other words, 
each seller can supply at most $q_i$ units of a homogeneous item. We assume that the sellers are strategic with respect to 
valuation per unit only and they always report their capacities truthfully. Let $x_i$ and $p_i$ denote the allocation and 
per unit payment to bidder $i$. Then the profit of the auction (or auctioneer) for bid vector $\textbf{b}$ is
 \begin{center}
   $\A(\textbf{b}, \C) = \C(\displaystyle \sum_{i=1}^{n}{x_i(\textbf{b}, \C)}) - 
   \displaystyle \sum_{i=1}^{n}{p_i(\textbf{b}, \C)\cdot x_i(\textbf{b}, \C)}$.
 \end{center}
The auctioneer wants to maximize her profit satisfying feasibility, IR, and DSIC. As before, we first define the notion
 of bid-independent auction for this setting.

\subsubsection*{Bid-independent Auction.}
 For each bidder $i$, the allocation and payments are determined in two phases as follows.
 \begin{enumerate} 
 \item Phase I :
 \vspace{-1mm}
 \begin{enumerate}
  \item $t_i \leftarrow f(\textbf{b}_{-i}, \C)$.
  \item If $t_i < v_i$, set $x_i \leftarrow 0$, $p_i \leftarrow 0$, and remove bidder $i$.
  \item Let $n^\prime$ be the number of remaining bidders.
 \end{enumerate} 
 \item Phase II : 
  \vspace{-2mm}
 \begin{enumerate}
  \item $i^\prime \leftarrow \underset{0 \leq i \leq m^\prime}{\operatorname{argmax}}\  \left(\C(i) - \displaystyle\sum_{j=1}^{k-1}{q_{[j]}t_{[j]}} - (i - \displaystyle\sum_{j=1}^{k-1}{q_{[j]}})t_{[k]}\right)$
   \\where $\ m^\prime = \sum_{j=1}^{n^\prime}{q_{[j]}}$ and $\sum_{j=1}^{k-1}{q_{[j]}} < i \leq \sum_{j=1}^{k}{q_{[j]}}$
  \item Suppose $k^\prime$ satisfies $\sum_{j=1}^{k^\prime-1}{q_{[j]}} < i^\prime \leq \sum_{j=1}^{k^\prime}{q_{[j]}}$.
  \item Set $x_{[i]} \leftarrow q_{[i]}$ and $p_{[i]} \leftarrow t_{[i]}$ for $i = \{1, \ldots, k^\prime - 1\}$.
  \item Set $x_{[k^\prime]} \leftarrow (i^\prime - \sum_{j=1}^{k^\prime-1}{q_{[j]}})$ and $p_{[k^\prime]} \leftarrow t_{[k^\prime]}$.
  \item Otherwise, set $x_{[i]} \leftarrow 0, p_{[i]} \leftarrow 0$.
 \end{enumerate}
 \end{enumerate} 
 As the allocation is monotone in bids and payment is bid-independent, any bid-independent auction is truthful.
 
\subsection{Prior-Free Benchmark}
We denote the $i$-th lowest valuation by $v_{[i]}$ and the corresponding capacity by $q_{[i]}$. Suppose $m = \sum_{i=1}^{n}{q_i}$.
Then we have, 
\begin{center}
  $\F(\textbf{b}, \C) = \max \limits_{0 \leq i \leq m} (\C(i) - iv_{[j]})$, where $\sum_{k=1}^{j-1}{q_{[k]}} < i \leq \sum_{k=1}^{j}{q_{[k]}}$
  $OPP(\textbf{b}, \C) = \underset{v_{[j]}}{\operatorname{argmax}}\left( \underset{\sum_{k=1}^{j-1}{q_{[k]}} < i \leq \sum_{k=1}^{j}{q_{[k]}}}\max (\C(i) - iv_{[j]}) \right)$
\end{center}
The first $j$ bidders are the winners and they are allocated at their full capacity except possibly the last one.
As no truthful auction can be constant-competitive against $\F$ we define $\F^{(2)}$ as the optimal single price auction that buys from 
at least two bidders. The profit of $\F^{(2)}$ on input vector \textbf{b} is
\begin{center}
  $\F^{(2)}(\textbf{b}, \C) = \max \limits_{q_{[1]} < i \leq m} (\C(i) - iv_{[j]})$, where $\sum_{k=1}^{j-1}{q_{[k]}} < i \leq \sum_{k=1}^{j}{q_{[k]}}$
\end{center}

\subsection{Profit Extracting Procurement Auction with Capacity (PEPAC)}
Now we extend the random sampling based procurement auction presented in Section \ref{rspa} for this setting.
\subsubsection*{Profit Extraction with Capacity (PEC$_P(\textbf{b}, \C)$).}
 \begin{enumerate}
  \item Find the largest value of $k^\prime$ for which $v_{[k]}$ is at most $(\C(k^\prime) - P)/k^\prime$ 
  where $\sum_{i=1}^{k-1}{q_{[i]}} < k^\prime \leq \sum_{i=1}^{k}{q_{[i]}}$.
  \item Pay these $k$ bidders $(\C(k^\prime) - P)/k^\prime$ per unit and reject others.
 \end{enumerate}

\subsubsection*{Profit Extracting Procurement Auction with Capacity (PEPAC).}
PEPAC is same as PEPA except that it invokes PEC$_P(\textbf{b}, \C)$ instead of PE$_P(\textbf{b}, \C)$.
Next we derive the performance of PEPAC through the following theorems. 
\begin{theorem}
\label{pepac1}
 PEPAC is $4$-competitive for any concave revenue curve $\C$ if \\$q_i = q \ \forall \ i \in \{1, \ldots, n\}$.
\end{theorem}
\begin{pfof}
 By definition, $\F^{(2)}$ on \textbf{b} buys from $k \geq 2$ bidders for a profit of $\F^{(2)}(\textbf{b}, \C) = \C(k^\prime) - k^\prime v_{[k]}$ 
 where $\sum_{i=1}^{k-1}{q} < k^\prime \leq \sum_{i=1}^{k}{q}$
 These $k$ bidders are divided uniformly at random between \y \ and \z. Let $k_1$ be the number of them in \y \ and $k_2$ 
 the number in \z. Now we denote the $i^{th}$ lowest bid in \y \ by $v_{[i]}^\prime$ and in \z \ by $v_{[i]}^{\prime\prime}$. Clearly, 
 $v_{[k_1]}^\prime \leq v_{[k]}$ and $v_{[k_2]}^{\prime\prime} \leq v_{[k]}$.
 As $\F^\prime$ and $\F^{\prime\prime}$ are optimal in respective partitions we have, 
 \begin{align*}
  \F^{\prime} \geq \C(k_1q) - k_1qv_{[k_1]}^{\prime} \geq \C(k_1q) - k_1qv_{[k]}.\\
  \F^{\prime\prime} \geq \C(k_2q) - k_2qv_{[k_2]}^{\prime\prime} \geq \C(k_2q) - k_2qv_{[k]}.
 \end{align*}
 
 Auction profit is $P =\ \min(F^\prime, F^{\prime\prime})$.
 Therefore,
 \begin{align*}
  \displaystyle \frac{\min(F^\prime, F^{\prime\prime})}{\F^{(2)}(\textbf{b}, \C)} &\geq 
  \frac{\min(\C(k_1q) - k_1qv_{[k]}, \C(k_2q) - k_2qv_{[k]})}{\C(k^\prime) - k^\prime v_{[k]}} \\
  &= \frac{\min(k_1q(\frac{\C(k_1q)}{k_1q} - v_{[k]}), k_2q(\frac{\C(k_2q)}{k_2q} - v_{[k]}))}{k^\prime(\frac{\C(k^\prime)}{k^\prime} - v_{[k]})} \\
  &\geq \frac{\min(k_1q(\frac{\C(k^\prime)}{k^\prime} - v_{[k]}), k_2q(\frac{\C(k^\prime)}{k^\prime} - v_{[k]}))}{k^\prime(\frac{\C(k^\prime)}{k^\prime} - v_{[k]})} \\
  & \displaystyle[\mbox{as} \  \frac{\C(k_1q)}{k_1q} \geq \frac{\C(k^\prime)}{k^\prime} \ \mbox{and}\  \frac{\C(k_2q)}{k_2q} \geq \frac{\C(k^\prime)}{k^\prime}\displaystyle]\\
  &= \frac{\min(k_1q, k_2q)}{k^\prime} \geq \frac{\min(k_1q, k_2q)}{kq} \\
  &= \frac{\min(k_1, k_2)}{k}.  
 \end{align*}
 
 Thus, the competitive ratio is given by 
 \begin{align*}
  \displaystyle \frac{\E[P]}{\F^{(2)}} &= \frac{1}{k} \sum_{i=1}^{k-1}{\min(i, k-i){k \choose i} 2^{-k}}\\
  &= \displaystyle \frac{1}{2} - {k-1 \choose \lfloor{k/2}\rfloor} 2^{-k}.\\
 \end{align*} 
 which is the same as in Theorem \ref{thm1}.
 \end{pfof}
\vspace{-2mm}
\begin{theorem}
\label{pepac2}
 PEPAC is $4 \cdot \left( \displaystyle \frac{q_{\max}}{q_{\min}}\right)$-competitive if the revenue curve is linear and 
 $q_i \in [q_{\min}, q_{\max}] \ \ \forall \ i \in \{1, \ldots, n\}$ and further this bound is tight.
\end{theorem}
\begin{pfof}
 By definition, $\F^{(2)}$ on \textbf{b} buys from $k \geq 2$ bidders for a profit of $\F^{(2)}(\textbf{b}, \C) = \C(k^\prime) - k^\prime v_{[k]}$ 
 where $\sum_{i=1}^{k-1}{q_{[i]}} < k^\prime \leq \sum_{i=1}^{k}{q_{[i]}}$
 These $k$ bidders are divided uniformly at random between \y \ and \z. Let $k_1$ be the number of bidders in \y \ and $k_2$ 
 the number in \z. Now we denote the $i^{th}$ lowest bid (according to valuation) in \y \ by $(v_{[i]}^\prime, q_{[i]}^\prime)$ and 
 in \z \ by $(v_{[i]}^{\prime\prime}, q_{[i]}^{\prime\prime})$. 
 Clearly, $v_{[k_1]}^\prime \leq v_{[k]}$ and $v_{[k_2]}^{\prime\prime} \leq v_{[k]}$.
 
 As $\F^\prime$ and $\F^{\prime\prime}$ are optimal in respective partitions we have, 
 \begin{align*}
  \F^\prime 
  \geq \C(\sum_{i=1}^{k_1}{q_{[i]}^\prime}) - (\sum_{i=1}^{k_1}{q_{[i]}^\prime})v_{[k]}, \ \ 
  \F^{\prime\prime} 
  \geq \C(\sum_{i=1}^{k_2}{q_{[i]}^{\prime\prime}}) - (\sum_{i=1}^{k_2}{q_{[i]}^{\prime\prime}})v_{[k]}.
 \end{align*}
 
 Suppose $\displaystyle\sum_{i=1}^{k_1}{q_{[i]}^\prime} = q_{x}$, 
 $\displaystyle\sum_{i=1}^{k_2}{q_{[i]}^{\prime\prime}} = q_{y}$, and $\displaystyle\sum_{i=1}^{k}{q_{[i]}} = q_{z}$.
 
 \begin{align*}
  \displaystyle \frac{\min(F^\prime, F^{\prime\prime})}{\F^{(2)}(\textbf{b}, \C)} &\geq 
  \frac{\min(\C(q_{x}) - q_{x}v_{[k]}, \C(q_{y}) - q_{y}v_{[k]})}{\C(k^\prime) - k^\prime v_{[k]}} \\
  &= \frac{\min(r q_{x} - q_{x}v_{[k]}, r q_{y} - q_{y}v_{[k]})}{r k^\prime - k^\prime v_{[k]}} \\
  &= \frac{\min(q_x, q_y)}{k^\prime} \geq \frac{\min(q_x, q_y)}{q_z} \\
  &\geq \frac{\min(k_1 q_{\min}, k_2 q_{\min})}{k q_{\max}} \\
  &= \left(\frac{q_{\min}}{q_{\max}}\right) \frac{\min(k_1, k_2)}{k} \geq \frac{q_{\min}}{4 \cdot q_{\max}}.
 \end{align*} 
 \end{pfof}
 \vspace{-2mm}
\begin{theorem}
\label{pepac3}
 PEPAC is $4 \cdot \left( \displaystyle \frac{q_{\max}}{q_{\min}}\right)$-competitive for any concave revenue curve $\C$ 
 if $q_i \in [q_{\min}, q_{\max}] \ \ \forall \ i \in \{1, \ldots, n\}$.
\end{theorem}

\section{Sellers with Non-Unit Strategic Capacities}
\label{strategic}
\subsection{Setup}
In this case, seller $i$ can misreport his capacity $q_i$ in addition to misreporting his valuation per unit $v_i$ to maximize his gain
from the auction. Here, we assume that sellers are not allowed to overbid their capacity. 
This can be enforced by declaring, as part of the auction, that if a seller fails to provide the 
number of units he has bid, he suffers a huge penalty (financial or legal loss). 
But underbidding may help a seller as depending on the mechanism it may result in an increase in the payment which can 
often more than compensate the loss due to a decrease in allocation.
Hence, as shown by Iyengar and Kumar \cite{iyengar2008optimal}, 
even when the bidders can only underbid their capacities, an auction that simply ignores the capacities of the bidders 
need not be incentive compatible. A small example can be constructed as follows. 

\begin{example}
Suppose the $(v_i, q_i)$ values of the sellers are $(6, 100), (8, 100), (10, 200)$ and $(12, 100)$. Consider an external market with 
maximum demand of $200$ units. The revenue curve is given by $R(j) = 15j$ when $j <= 200$ and $15 * 200$ when $j > 200$. Suppose the buyer 
conducts the classic Kth price auction where the payment to a winning seller is equal to the valuation of the first losing seller. 
Bidding valuation truthfully is a weekly 
dominant strategy of the sellers but it does not deter them from possibly altering their capacities. If they report both $v_i$ and $q_i$ 
truthfully the allocation will be $(100, 100, 0 , 0)$ and the utility of the second seller will be $(10-8) * 100 = 200$. If the second seller 
underbids his capacity to $90$ the allocation changes to $(100, 90, 10, 0)$ and the utility of the second seller will be $(12-8) * 90 = 360$. 
So the Kth price auction is clearly not incentive compatible. 
\end{example}

Note that the actual values of $b_i = (v_i,q_i)$ are known to only seller $i$. From now on, the true type of each bidder is 
represented by $b_i = (v_i,q_i)$ and each reported bid is represented by $\hat{b}_i = (\hat{v}_i,\hat{q}_i)$. So we denote the true 
type vector by \textbf{b} and the reported bid vector by $\hat{\textbf{b}}$. 
We denote the utility or \emph{true} surplus of bidder $i$ by $u_i(\hat{\textbf{b}}, \C)$ and the \emph{offered} surplus by 
$\hat{u}_i(\hat{\textbf{b}}, \C)$. True surplus is the pay-off computed using the true valuation and 
the offered surplus is the pay-off computed using the reported valuation. 
\begin{center}
 $u_i(\hat{\textbf{b}}, \C) = [p_i(\hat{\textbf{b}}, \C)x_i(\hat{\textbf{b}}, \C) - v_ix_i(\hat{\textbf{b}}, \C)]$.
 $\hat{u}_i(\hat{\textbf{b}}, \C) = [p_i(\hat{\textbf{b}}, \C)x_i(\hat{\textbf{b}}, \C) - \hat{v}_ix_i(\hat{\textbf{b}}, \C)]$.
\end{center}
  
\subsection{A Characterization of DSIC and IR Procurement Auctions}
Iyengar and Kumar \cite{iyengar2008optimal} have characterized all DSIC and IR procurement auctions and 
 the payment rule that implements a given DSIC allocation rule.
\label{ch}
\begin{enumerate}
 \item \label{c1} A feasible allocation rule \textbf{x} is \textbf{DSIC} if, and only if, $x_i(((v_i,q_i),\hat{b}_{-i}), \C)$ is non-increasing in $v_i$,$\ \forall q_i $, $ \forall \hat{b}_{-i}$.
 \item \label{c2} A mechanism (\textbf{x}, \textbf{p}) is \textbf{DSIC} and \textbf{IR} if, and only if, the allocation rule \textbf{x} 
 satisfies \ref{c1} and the ex-post offered surplus is 
 \begin{center}
  $\hat{u}_i(\hat{b}_i, \hat{b}_{-i}, \C) = \displaystyle \int_{\hat{v}_i}^{\infty} x_i(u, \hat{q}_i, \hat{b}_{-i}, \C)\mathrm{d}u$.
 \end{center}
 with $\hat{u}_i((\hat{v}_i, \hat{q}_i), \hat{b}_{-i}, \C)$ non-negative and non-decreasing in $\hat{q}_i$ for all $\hat{v}_i$ and for all $\hat{b}_{-i}$. 
\end{enumerate}

 We use the same prior-free benchmark and the random sampling procurement auction as defined in Section \ref{Nonstrategic} 
 and extend our previous results for strategic capacity case.
 
 \begin{lemma}
  \label{pesc1}
  PEC$_P$ is truthful if $\C$ is linear.
 \end{lemma}
 \begin{pfof}
 Let $k_1$ and $k_2$ be the number of units PEC$_P$ procures from the bidders to achieve target profit $P$ when bidders report their 
 capacities truthfully and misreport their capacities respectively. By our assumption bidders are only allowed to underbid their 
 capacities. So we have $k_2 \leq k_1$. Now truthful reporting is a dominant strategy of the bidders if the bidders are not better off 
 by underbidding their capacities. So $\hat{q}_i \leq q_i$ and 
  $u_i(((v_i,\hat{q}_i),\hat{b}_{-i}), \C) \leq u_i(((v_i,q_i),\hat{b}_{-i}), \C)$. Hence we have,
 \begin{center}
  $\displaystyle -v_i\hat{q}_i + \hat{q}_i\frac{\C(k_2) - P}{k_2} \leq  -v_iq_i + q_i\frac{\C(k_1) - P}{k_1}$.
 \end{center}
 
 A sufficient condition for the above inequality to hold is
 \begin{center}
 $\displaystyle \frac{\C(k_2) - P}{k_2} \leq \frac{\C(k_1) - P}{k_1} \ \ \forall k_2 \leq k_1$.  
 \end{center}
 Clearly a linear revenue curve satisfies the above sufficient condition. 
 \end{pfof}
 
 The following results immediately follow from the above lemma.
 \begin{theorem}
  \label{pepasc1}
  PEPAC is truthful if $\C$ is linear.
 \end{theorem}
\begin{theorem}
 PEPAC is $4$-competitive if the revenue curve $\C$ is linear and \\$q_i = q \ \forall \ i = \{1, \ldots, n\}$.
\end{theorem}
\begin{theorem}
 PEPAC is $4 \cdot \left( \displaystyle \frac{q_{\max}}{q_{\min}}\right)$-competitive for any linear revenue curve $\C$ 
 if $q_i \in [q_{\min}, q_{\max}] \ \forall i \in \{1, \ldots, n\}$.
\end{theorem}

 \section{Conclusion and Future Work}
 \label{future}
 In this paper, we have considered a model of prior-free profit-maximizing procurement auctions with capacitated sellers and designed 
 prior-free auctions for both single dimensional and bi-dimensional sellers. We have shown that the optimal single price auction 
 cannot be matched by any truthful auction. Hence, we have considered a lightly constrained single price auction as our benchmark in the analysis.
 We have presented procurement auctions based on profit 
 extraction, PEPA for sellers with unit capacity and PEPAC for sellers with non-unit capacity and proved an upper bound on their competitive ratios. 
 For the bi-dimensional case, PEPAC is truthful for the specific case of linear revenue curves.
 Our major future work is to design a prior-free auction for bi-dimensional sellers which is truthful and competitive 
 for all concave revenue curves.
 Subsequently, we would like to design prior-free procurement auctions for the more generic setting 
 where each seller can announce discounts based on the volume of supply.  
 
\newpage 
\bibliography{ref}

\end{document}